\documentclass[%
 aip,
 amsmath,amssymb,
reprint,%
]{revtex4-1}

\usepackage{graphicx}
\usepackage{bm}

\usepackage[utf8]{inputenc}
\usepackage[T1]{fontenc}
\usepackage{mathptmx}
\usepackage{etoolbox}
\usepackage{color}

\makeatletter
\def\@email#1#2{%
 \endgroup
 \patchcmd{\titleblock@produce}
  {\frontmatter@RRAPformat}
  {\frontmatter@RRAPformat{\produce@RRAP{*#1\href{mailto:#2}{#2}}}\frontmatter@RRAPformat}
  {}{}
}
\makeatother
\begin{document}

\title{Gas-to-soliton transition of attractive bosons on a spherical surface}

\author{A. Tononi}
\affiliation{ICFO-Institut de Ciencies Fotoniques, The Barcelona Institute of Science and Technology, Av. Carl Friedrich Gauss 3, 08860 Castelldefels, Barcelona, Spain}
\affiliation{Universit\'e Paris-Saclay, CNRS, LPTMS, 91405 Orsay, France}

\author{G. E. Astrakharchik}
\affiliation{Department de F\'isica, Universitat Polit\`ecnica de Catalunya, Campus Nord B4-B5, E-08034, Barcelona, Spain}

\author{D. S. Petrov}
\affiliation{Universit\'e Paris-Saclay, CNRS, LPTMS, 91405 Orsay, France}

\date{\today}

\begin{abstract}
We investigate the ground state properties of $N$ bosons with attractive zero-range interactions characterized by the scattering length $a>0$ and confined to the surface of a sphere of radius $R$. We present the analytic solution of the problem for $N=2$, mean-field analysis for $N\rightarrow \infty$, and exact diffusion Monte-Carlo results for intermediate $N$. For finite $N$, we observe a smooth crossover from the uniform state in the limit $a/R\gg 1$ (weak attraction) to a localized state at small $a/R$ (strong attraction). With increasing $N$ this crossover narrows down to a discontinuous transition from the uniform state to a soliton of size $\sim R/\sqrt{N}$. The two states are separated by an energy barrier, tunneling under which is exponentially suppressed at large $N$. The system behavior is marked by a peculiar competition between space-curvature effects and beyond-mean-field terms, both breaking the scaling invariance of a two-dimensional mean-field theory.
\end{abstract}

\maketitle

\section{\label{sec:level1}Introduction}

In two dimensions, $N$ bosonic atoms with attractive point-like interactions form bound states with very peculiar properties. Addressing this problem by using the mean-field (MF) density functional theory characterized by the coupling constant $g$ and normalization $N$ leads to a solution called Townes soliton, first studied in nonlinear optics~[\onlinecite{chiao1964}] and recently observed in cold-atom experiments~[\onlinecite{bakkalihassani2021},\onlinecite{chen2021}]. This state exists only for a specific value of $g=g_c=-4\pi\hbar^2/(mN\ln r)$, with $r=8.567$; a stronger (weaker) attraction yields a collapsing (expanding) state~[\onlinecite{chiao1964}]. For $g=g_c$, the shape of the soliton is defined only up to an arbitrary scaling factor and the total MF energy vanishes independent of the soliton size~[\onlinecite{Vlasov,Pitaevskii,PitaevskiiRosch,Dalibard2022}].

Hammer and Son~[\onlinecite{hammer2004}] have argued that the two-dimensional coupling constant depends logarithmically on the typical length, which breaks the MF degeneracy and fixes a specific soliton size. They predict that in the limit of large $N$, the soliton shrinks by $\sqrt{r}=2.927$ every time one adds another atom while the energy $B_N$ scales as $B_N/B_2\propto r^N$. Here $B_2=-4\hbar^2e^{-2\gamma}/(ma^2)$ is the dimer energy and $\gamma$ is the Euler constant. These predictions complement the exact few-body numerics performed over a few decades for $N \le 4$~[\onlinecite{bruch1979, adhikari1988, nielsen1997, nielsen1999, brodsky2006, kartavtsev2006}]. Bazak and Petrov~[\onlinecite{bazak2018}] have calculated $B_N$ for $N$ up to 26.

In this paper, we consider the problem of $N$ bosons confined on a spherical surface and interacting via zero-range attraction. In the MF approximation on a flat surface there is a critical coupling constant $g_c$, which separates the collapsed ($g<g_c$) and uniform ($g>g_c$) ground states. However, the curved geometry breaks the MF scaling invariance. As a result, the uniform state remains metastable in a finite interval $g_d<g<g_c$, where the uniform and collapsed states are separated by an energy barrier. These are characteristics of a first-order phase transition, the system size taken as the order parameter and the point $g=g_d=-2\hbar^2\pi/(mN)$ being the spinodal on the uniform phase side. Here, the barrier and the frequency of the lowest (dipole) excitation of the uniform condensate vanish. These results are obtained in the $N \to \infty$ limit, for which the classical Gross-Pitaevskii description is exact. However, due to the singular behavior of the collapsed state { and the fact that it does not constitute a good starting point for a $1/N\ll 1$ expansion,} the problem becomes essentially quantum as soon as one passes from $N=\infty$ to finite $N$. We argue that in this case the discontinuous gas-collapse transition turns into an avoided crossing between the uniform state and the Townes soliton of size $\sim R/\sqrt{N}$, the tunneling between these states being exponentially suppressed at large $N$. To confirm these predictions we solve the two-body problem analytically{, perturbatively analyze the gas and the soliton solutions,} and perform diffusion Monte-Carlo (DMC) and variational Monte-Carlo (VMC) analysis up to $N=128$. 

Our results indicate that the change from the smooth crossover to discontinuous transition (for all practical purposes) happens approximately at $N\sim 10$, which is quantitatively large. We attribute this effect to the relatively weak breaking of the scaling invariance, which results in a quantitatively low energy barrier between the phases. This means that linear superpositions of the uniform and soliton states and macroscopic tunneling dynamics could be observable in this system for rather large atom numbers $10\lesssim N\lesssim 20$.

Note that the uniform-to-soliton transition for bosons on a sphere is very different from what happens with their one-dimensional counterpart. Indeed, for attractive one-dimensional bosons on a ring this transition is continuous in both infinite-$N$ and finite-$N$ cases~[\onlinecite{Lincoln},\onlinecite{kanamoto2003}]. 

Our study is inspired by the growing experimental interest in shell-shaped bosonic gases~[\onlinecite{carollo2022},\onlinecite{jia2022},\onlinecite{guo2022}] and by the goal of producing low-dimensional atomic devices whose curved geometry acts as a tunable degree of freedom for quantum simulation~[\onlinecite{tononi2023},\onlinecite{amico2022}].

\section{Mean-field analysis in the large-$N$ limit}
\label{Sec:MFbasics}

We start our discussion with the MF analysis. Setting $\hbar=m=R=1$ the Hamiltonian of the system reads
\begin{equation}\label{Ham}
\hat{H} = \frac{1}{2}\int d\Omega (\hat{\psi}_{\Omega}^{\dagger} \hat{L}^2 \hat{\psi}_{\Omega} + g \hat{\psi}_{\Omega}^{\dagger} \hat{\psi}_{\Omega}^{\dagger} \hat{\psi}_{\Omega} \hat{\psi}_{\Omega}), 
\end{equation}
where $\hat{\psi}_{\Omega}^{\dagger}$ is the operator creating a boson at position $\Omega=(\theta,\varphi)$ on the sphere, $g$ is the coupling constant, and $\hat{L}^2=-[\partial_{\theta}^2+\cot \theta \, \partial_{\theta}+(1/\sin^2\theta)\partial^2_\varphi]$. To regularize the local interaction term in Eq.~(\ref{Ham}) we use a cutoff for the interaction at an angular momentum $l_c\gg 1$ (equivalent to the momentum cutoff at $\kappa=l_c$). The coupling constant is then taken as 
\begin{equation}\label{gkappa}
g=4\pi/\ln(|B_2|/\kappa^2)=2\pi/\ln[2e^{-\gamma}/(\kappa a)].
\end{equation}

On the MF level we approximate $\hat{\psi}_{\Omega}^{\dagger}=\hat{\psi}_{\Omega}=\Phi(\Omega)$ and minimize Eq.~(\ref{Ham}) with the normalization constraint $\int d\Omega |\Phi(\Omega)|^2=N$. In this manner, if we also assume the cylindrical symmetry (no dependence on the azimuthal angle $\varphi$), we arrive at the Gross-Pitaevskii equation
\begin{equation}\label{GPEsphere}
\left[-\frac{1}{2} (\partial_{\theta}^2
+\cot \theta \, \partial_{\theta})
+ g |\Phi(\theta)|^2 -\mu \right] \Phi(\theta) = 0.
\end{equation}
The first obvious solution of Eq.~(\ref{GPEsphere}) is the uniform condensate $\Phi=\sqrt{N/(4\pi)}$ corresponding to the energy $E_N=gN^2/(8\pi)$. The spectrum of the Bogoliubov modes in this case reads $\epsilon_{l,m}=\sqrt{l(l+1)[l(l+1)+gN/\pi]}/2$~[\onlinecite{prestipino2019},\onlinecite{tononi2019}]. The uniform solution is thus stable with respect to small amplitude modulations for $g$ larger than $g_d=-2\pi/N$. Just below this point the lowest-lying dipole mode ($l=1$) becomes unstable. On the other hand, since $g_d<g_c$, in the interval $g_d<g<g_c$ the uniform state can only be metastable. Indeed, one can consider the Townes soliton solution with the angular width $\delta \theta$ and show that its variational energy diverges as $\sim (g-g_c)/\delta\theta^2$ in the asymptotic limit $\delta \theta\rightarrow 0$. 

The conclusion that the ground state is, respectively, the collapsed state for $g<g_c$ and the uniform (gas) state for $g>g_c$, is confirmed by our numerical analysis of Eq.~(\ref{GPEsphere}) based on imaginary-time propagation. In practice, we see the collapsed state as a solution localized on a few sites of our spatial grid. This solution is nonuniversal { since its size is determined by the finite grid spacing and it shrinks to a delta-function distribution when increasing the density of the mesh.} The imaginary-time propagation method does not predict other nonuniform solutions. On the other hand, by using a shooting method{, which can in principle also describe maxima or saddle points of the energy functional,} we do observe a universal (independent of the grid) nonuniform nodeless solution in the interval $g_d<g<g_c$. We identify it as a saddle point separating the uniform state from the collapsed one. Its energy is always larger than the energy of the uniform state. The maximal energy barrier $\Delta E_{\rm max}\approx 0.0336 N$ is attained at $g=g_c$. The gas-soliton transition at $g_c$ is thus discontinuous. The barrier decreases as $g$ approaches $g_d$ and disappears completely at $g_d$, which we identify as the spinodal point, consistent with the Bogoliubov analysis. We should note that the gas-soliton transition breaks the rotational symmetry on a sphere in the same manner as it breaks the translational symmetry in the flat case, and Eq.~(\ref{GPEsphere}) explicitly takes this symmetry breaking into account.

{ The MF (or classical) Gross-Pitaevskii equation is the leading-order description of a weakly interacting Bose condensate. It holds when the number of atoms per healing volume is large (see, for instance, Ref.~[\onlinecite{MoraCastin2003}]). In two dimensions, the healing volume scales as $\sim 1/|\mu|\sim 1/(|g||\Phi|^2)$ and therefore it contains $\sim 1/|g|$ atoms. The MF theory is thus valid when $|g|\ll 1$. In our case this inequality is} equivalent to $N\gg 1$ since we are interested in the region close to $g_c=-4\pi/(N\ln r)=-1.862\pi/N$~{ [\onlinecite{ExpA}]}. A conceptual problem when going from $N=\infty$ to finite $N$ is caused by the singular MF description of the soliton state. Its size vanishes and the binding energy diverges for $g<g_c${, which is not a good starting point for a perturbation theory}. This difficulty is resolved by the beyond-MF analysis of Hammer and Son~[\onlinecite{hammer2004}] who argue that the soliton has a finite size proportional to $a$ (although with a prefactor exponentially small for large $N$) and finite energy $\propto 1/a^2$. Compared to the energy of the uniform state, proportional to $g$, the soliton energy varies exponentially fast, $\propto 1/a^2\propto e^{4\pi/g}$. We thus conjecture that for large but finite $N$ the gas-soliton transition is a crossing of these two states. Although not directly applicable to describe the crossing, the MF theory allows us to make two important statements. First, the MF description suggests that the barrier separating the two phases grows linearly with $N$, the crossing narrows down exponentially with $N$. Second, the MF results also suggest that for sufficiently large $N$ the soliton at the crossing is small { and is only weakly influenced by the local curvature. In the following Secs.~\ref{Sec:2body} and \ref{Sec:Soliton} we will use the inequality $N\gg 1$ to perturbatively describe, respectively, the uniform solution and the soliton solution. The transition point can then be located by comparing the corresponding energies. In Sec.~\ref{Sec:MC} we will present our numerical results obtained for finite $N$.}

\section{Renormalization of the interaction and the two-body problem}
\label{Sec:2body}

\begin{figure}
\centering
\includegraphics[width=0.99\columnwidth]{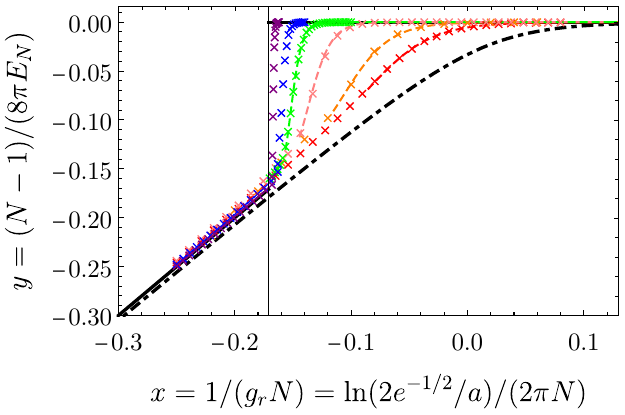}
\caption{Inverse of the energy along the crossover from the weakly attractive (large negative $x$) to strongly attractive (large positive $x$) regimes. The cases of $N=2$ and $N=\infty$ are shown as the black dashed-dotted curve and the piecewise linear black solid curve, respectively. The transition in the thermodynamic limit takes place at $x=-0.171$. The red, orange, pink, green, blue, and purple crosses are the Monte Carlo results, respectively, for $N=3,4,8,16,32$, and 128 bosons. The dashed curves { represent the large-$x$ asymptote Eq.~(\ref{flatxy}), which takes into account the leading-order curvature-induced energy shift for a nearly flat soliton.} 
}
\label{fig1}
\end{figure}

In contrast to the soliton state, the uniform solution profits from a well-behaved MF description. Therefore, the MF formula $E_N=gN^2/(8\pi)$ is sufficient for the asymptotic large-$N$ analysis. Nevertheless, we would like to renormalize $g$ and express it in a cutoff independent form. One way of performing this task is to note that the energy of a Bose gas in the weakly-interacting limit is the interaction energy shift for a single pair multiplied by the number of pairs. In this approach the renormalized $g/(4\pi)$ can be identified with the energy $E_2$ of a pair of atoms, which can be calculated directly from the two-body Schrödinger equation for the relative motion
\begin{equation}\label{Schr}
\hat{L}^2 \psi(\theta) =  E_2 \, \psi(\theta).
\end{equation}
The relative wave function $\psi(\theta)$ is regular everywhere except $\theta\rightarrow 0$ where it satisfies the Bethe-Peierls boundary condition $\psi(\theta)\propto \ln(\theta/a)$. The suitable solution is the associated Legendre function $\psi(\theta)\propto P_{-1/2+s}[\cos(\pi-\theta)]$ and the parameter $s=\sqrt{E_2+1/4}$ satisfies 
\begin{equation}\label{twobodyenergy}
\ln(1/a) = [\Psi_0(1/2+s) + \Psi_0(1/2-s)]/2 +\ln(e^{\gamma}/2),
\end{equation}
where $\Psi_0$ is the digamma function. Equation~(\ref{twobodyenergy}) implicitly defines the function $E_2(a)$, which we find to smoothly connect the limits of strong ($a\ll 1$) and weak ($a\gg 1$) interactions (see Fig.~\ref{fig1}). In the strongly interacting limit we { obtain the expansion $E_2=-4\exp(-2\gamma)/a^2-1/3+o(1)$, and, therefore,} reproduce the flat-surface asymptotics $E_2\approx B_2$. For discussing the weakly interacting limit $a\gg 1$ let us introduce
\begin{equation}\label{gr}
g_r=\frac{2\pi}{\ln(2e^{-1/2}/a)}
\end{equation} 
and expand Eq.~(\ref{twobodyenergy}) for small $|E_2|$. In this manner we obtain the series
\begin{equation}\label{E2Pert}
E_2=\frac{g_r}{4\pi}-\left(\frac{g_r}{4\pi}\right)^3+ o(g_r^3).
\end{equation}
Thus, $g_r$ can be identified as the renormalized coupling constant suitable for describing the two-dimensional scattering in our particular geometry. 
{ Note that the energy can be equivalently expanded either in powers of $g$ or in powers of $g_r$, the corresponding series deviate from each other at the beyond-MF level since $g-g_r\sim g^2\ln \kappa \ll |g|$.} We also note that the second-order term in Eq.~(\ref{E2Pert}) vanishes because of our choice of the constant under the logarithm in Eq.~(\ref{gr}).

Equation~(\ref{E2Pert}) is the energy of two atoms calculated up to terms $\propto g^3\sim g_r^3$. One can also generalize it to arbitrary $N$ by using the standard perturbation expansion of the Hamiltonian~(\ref{Ham}) at small $g$. Up to the third order this procedure gives (see the derivation in Appendix~\ref{appendix}, based on Ref.~[\onlinecite{PricoupenkoPetrov2021}])
\begin{equation}\label{EnUniformRen}
E_N=\left[\frac{g_r}{4\pi}-\left(\frac{g_r}{4\pi}\right)^3\right]\frac{N(N-1)}{2}+\left(\frac{g_r}{4\pi}\right)^3N(N-1)(N-2).
\end{equation}
As expected, Eq.~(\ref{EnUniformRen}) reproduces the perturbative two-body result Eq.~(\ref{E2Pert}) and also predicts the leading nonpairwise contribution to the energy in the weakly interacting limit. Note that the three-body and, more generally, other beyond-MF terms, are suppressed by at least one additional power of $g\propto 1/N$ compared to the leading two-body term $E_N=g_rN^2/(8\pi)$.

{
\section{Nearly flat soliton and the transition point}

\label{Sec:Soliton}

Let us assume (and {\it a posteriori} verify) that for large $N$ there is a localized soliton solution of Eq.~(\ref{Ham}), degenerate with the uniform solution at a certain critical $g_r$. By localized we mean that the soliton size is much smaller than the sphere radius. In this nearly flat limit we can address the problem perturbatively.

As a measure of the cloud size, it is convenient to introduce the rms separation between pairs of atoms $\sqrt{\langle r_{ij}^2 \rangle}$. In the spherical geometry, $\langle r_{ij}^2\rangle$ is calculated from the distribution of the three-dimensional (chord) distances $r_{ij}=|{\bf r}_i-{\bf r}_j|$, where ${\bf r}_i$ is the three-dimensional coordinate corresponding to the point $\Omega_i$ on the sphere. 

As we mention in the introduction, in the limit $N\rightarrow \infty$, the size of the soliton on a plane equals~[\onlinecite{hammer2004}]
\begin{equation}\label{sigmaN}
\sqrt{\langle r_{ij}^2 \rangle}=ae^{-N(\ln r)/2+o(N)}
\end{equation}
and its energy (to avoid confusion, we always denote energies on a sphere by $E$ and on a plane by $B$)
\begin{equation}\label{BN}
B_N=-a^{-2}e^{N\ln r +o(N)}=-e^{N\ln r + 4\pi/g_r+o(N)},
\end{equation}
where the second equality in Eq.~(\ref{BN}) follows from the definition of $g_r$ Eq.~(\ref{gr}). One can show that 
\begin{equation}\label{BvsRresult}
\langle r_{ij}^2 \rangle |B_N|=0.553,
\end{equation}
which is a stronger statement than what can be obtained by simply multiplying Eqs.~(\ref{sigmaN}) and (\ref{BN})~[\onlinecite{footnoteoN}]. We emphasize that $B_N\sim 1/\langle r_{ij}^2 \rangle$ is a beyond-MF energy scale. It is indeed much smaller than the MF kinetic or interaction energies, which both scale as $\sim N/\langle r_{ij}^2 \rangle$, but have different signs and almost cancel each other. A detailed derivation of Eq.~(\ref{BvsRresult}) is presented in Appendix B where we also review some properties of flat solitons.

Passing now to the curved geometry one can show that the leading-order effect of the curved spherical surface on a nearly flat soliton is to shift its energy by $\Delta E_{\rm sphere}$ according to
\begin{equation}\label{DeltaEsphereGen}
E_N=B_N+\Delta E_{\rm sphere}=B_N-0.199N.
\end{equation}
To obtain Eq.~(\ref{DeltaEsphereGen}), we use the flat Townes soliton as the unperturbed solution and the difference between the kinetic energy operator on a sphere and on a plane as the perturbation (refer to Appendix~C for more details). It is important to emphasize that for the validity of Eq.~(\ref{DeltaEsphereGen}) the curvature shift $\Delta E_{\rm sphere}$ should be small compared to the MF energy scale $\sim N/\langle r_{ij}^2\rangle$ and not to the beyond-MF energy scale $B_N$. As a result, without invalidating Eq.~(\ref{DeltaEsphereGen}), both terms on its right-hand side can be comparable to each other. As we will now see, at the transition point, the curvature effect indeed competes with the beyond-MF effect.

Combining the results of Sec. \ref{Sec:2body} on the energy of the uniform phase and Eq.~(\ref{DeltaEsphereGen}) for the soliton we obtain the condition for their crossing (valid for large $N$ and $1/\langle r_{ij}^2\rangle$)
\begin{equation}\label{DetTransMainText}   
g_r N^2/(8\pi)+o(N)=B_N+\Delta E_{\rm sphere}+o(N,1/\langle r_{ij}^2 \rangle).
\end{equation}
Substituting Eq.~(\ref{BN}) into Eq.~(\ref{DetTransMainText}), we find that the critical coupling constant equals $g_r=-4\pi/(N\ln r)+o(N^{-1})\approx g_c$. This conclusion just follows from the fact that at the crossing $B_N$ given by Eq.~(\ref{BN}) cannot be exponentially large in $N$ as there are no other exponentially large terms in Eq.~(\ref{DetTransMainText}). We, thus, recover the MF result that the transition point corresponds to $g_r=g_c$.  Substituting this equality into the left-hand side of Eq.~(\ref{DetTransMainText}) tells us that at the crossing $B_N=-N/(2\ln r)+0.199 N$ and, using Eq.~(\ref{BvsRresult}), we obtain the following estimate for the critical soliton size
\begin{equation}\label{rmscrit}
\sqrt{\langle r_{ij}^2\rangle}|_{\rm crit}=4.06/\sqrt{N}.
\end{equation} 
In Eq.~(\ref{DetTransMainText}), we keep track of possible corrections; we include the beyond-MF correction $o(N)$ for the uniform phase on the left-hand side and an estimate of higher-order beyond-MF and finite-curvature corrections for the soliton on the right-hand side. Using Eq.~(\ref{rmscrit}), one can {\it a posteriori} verify that these corrections are small.

We have thus confirmed that for large $N$ the gas-soliton transition is a crossing of the uniform state which occupies the whole sphere surface and a soliton of size $\sim 1/\sqrt{N}$. In passing, we note that in order to observe a significant size change at the transition point the number of atoms should be $N\gtrsim 16$, as follows from Eq.~(\ref{rmscrit}). In Sec. \ref{Sec:Crossover} we discuss what happens with the gas-soliton transition for finite $N$.  

\section{Numerical results}
\label{Sec:Crossover}

In Fig.~\ref{fig1}, we show the energy obtained by solving the $N$-boson problem on a sphere by means of the diffusion and variational Monte Carlo methods, details of which will be discussed in Sec.~\ref{Sec:MC}. The dots in Fig.~\ref{fig1} stand for the DMC results for $N=3$ (red), $N=4$ (orange), $N=8$ (pink), $N=16$ (green), $N=32$ (blue), and $N=128$ (purple). The $N=2$ result Eq.~(\ref{twobodyenergy}) is plotted as the black dashed-dotted curve.

For presenting the data we choose to plot the quantity $y=(N-1)/(8\pi E_N)$ as a function of $x=1/(g_r N) = \ln(2e^{-1/2}/a)/(2\pi N)$. With this choice of coordinates, in the weakly interacting limit ($x\rightarrow -\infty$), the curves $y(x)$ for different $N$ should all tend to the straight black solid $y=x$ line, equivalent to $E_N=g_rN(N-1)/(8\pi)$. We note that the convergence of the finite-$N$ results to this line with increasing $N$ is not monotonic and is relatively slow, which is very well explained by the cubic terms in Eq.~(\ref{EnUniformRen}). We do not show the corresponding curves to avoid cluttering. 

In the opposite strongly interacting limit ($x\rightarrow +\infty$) our numerical results agree with Eq.~(\ref{DeltaEsphereGen}), which, in variables $x$ and $y$  explicitly reads
\begin{equation}\label{flatxy}
y= - \frac{N-1}{8 \pi } \frac{1}{(B_N/B_2)\exp(4\pi N x +1-2\gamma)-\Delta E_{\rm sphere}(N)}.
\end{equation}
In fact, we calculate the curvature induced offset $\Delta E_{\rm sphere}(N)$ for arbitrary $N$, not only for $N\gg 1$. However, we find that $\Delta E_{\rm sphere}(N)$ converges very fast to the large-$N$ asymptote Eq.~(\ref{DeltaEsphereGen}). For $N=3$, we numerically find $\Delta E_{\rm sphere}(3)=0.60(2)$, and the distinction is considerable only for $N=2$, where $\Delta E_{\rm sphere}(2)=-1/3$ (see Sec.~\ref{Sec:2body}).
The dashed curves in Fig.~\ref{fig1} show Eq.~(\ref{flatxy}) for $N=3$, 4, 8, and 16 using the ratios $B_N/B_2$ calculated in Ref.~[\onlinecite{bazak2018}]. 

We observe that with increasing $N$ the energy curves in Fig.~\ref{fig1} tend to the piecewise linear function, $y=x$ (gas phase) for $g_r<g_c$ and $y=0$ (collapsed state) for $g_r>g_c$, consistent with the MF description of Sec.~\ref{Sec:MFbasics}. The transition in the limit $N\rightarrow \infty$ is marked by the vertical line at $x=-\ln r/(4\pi) = -0.171$ (equivalent to $g_r=g_c$).
}

\begin{figure}
\centering
\includegraphics[width=0.99\columnwidth]{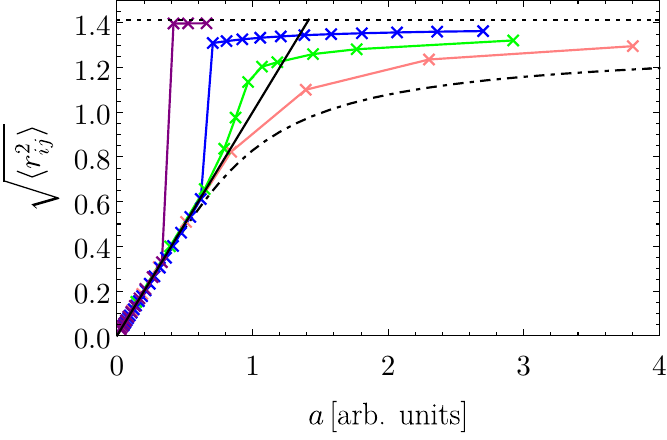}
\caption{The rms separation versus $a$ for different values of $N$ as in Fig.~\ref{fig1} (we use the same color code). The interatomic distance is defined as the three-dimensional chord length. The horizontal axis is proportional to the scattering length $a$ rescaled by an $N$-dependent coefficient to make all curves tend to a unique line at small $a$. The nonmonotonic behavior of the slope of the curves for $N\geq 16$ signals the abruptly changing size in the vicinity of the gas-soliton transition.
}
\label{fig2}
\end{figure}

Let us now discuss how the cloud size changes as we vary the interaction strength. In Fig.~\ref{fig2}, we show the rms separation between atoms as a function of the scattering length $a$ (rescaled by an $N$-dependent coefficient). In the strongly interacting limit, we are dealing with a localized and almost flat soliton, the size of which is proportional to $a$ { [see Eq.~(\ref{sigmaN})]}. Accordingly, we fit this linear dependence and rescale the horizontal axis to make the curves for different $N$ collapse to a single line. In the opposite weakly interacting limit, the distribution of atoms on our unit sphere is uniform and the rms size tends to $\sqrt{2}$. The way these two asymptotes are approached depends on $N$. We find that for $N\lesssim 10$ the derivatives of the curves are monotonic (within our precision) and for larger $N$ we start seeing a nonmonotonic structure, which eventually transforms into an abrupt change at a certain critical $a$. Our calculations are consistent with the scenario that there is a (narrow) crossing region where the exact ground state is a linear superposition of the soliton state and the uniform gas state. However, beginning with $N\sim 20$, our DMC scheme does not account for these effects. For $N=32$ and $128$ the DMC scheme predicts the energies and rms sizes of the two phases, but due to the importance sampling (necessary for these large $N$) the system gets stuck in one of the phases and cannot tunnel to the other. Neglecting this macroscopic tunneling, one can say that the system undergoes a first-order phase transition (see the discussion in Sec.~\ref{Sec:MC}). We find that for $N=32$ and, particularly, for $N=128$, the rms size of the soliton at the transition point is in quantitative agreement with Eq.~(\ref{rmscrit}).

\begin{figure}
\centering
\includegraphics[width=0.99\columnwidth]{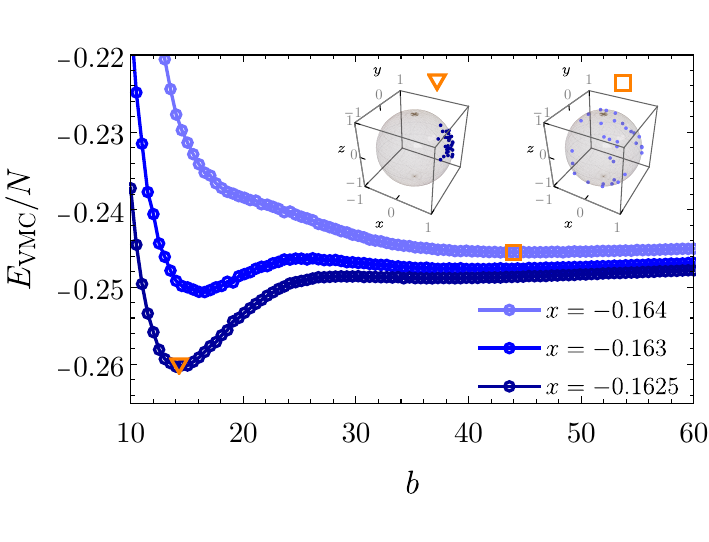}
\caption{
The variational energy per particle $E_{\text{VMC}}/N$ as a function of the localization parameter $b$ for $N=32$ and three close values of $x$ near the transition: for weak attraction (upper curve), there is a single minimum at large $b\gg 1$ physically corresponding to a uniform gas state, for strong attraction (bottom curve), the minimum is located at significantly smaller $b$ and describes a localized state, the middle curve shows the double minimum structure, typical for a first-order phase transition. The left and right insets show snapshots of particles' coordinates in VMC simulation in the minimum of the lower curve (marked by triangle, $b=14$) and in the minimum of the upper curve (square, $b=44$), respectively.
}
\label{fig3}
\end{figure}

\section{Monte Carlo for bosons on a sphere}
\label{Sec:MC}
To find the ground state of the $N$-boson problem numerically, we employ the variational (VMC) and diffusion (DMC) Monte Carlo methods, general details of which can be found, for instance, in Ref.~[\onlinecite{BoronatCasulleras1994}]. 
The standard method has to be adapted for solving the Schrödinger equation on a sphere.
As the variational (for VMC) and guiding (for DMC) wave function we take the Jastrow product 
\begin{equation}\label{Jastrow}
\Psi_0(\Omega_1,\cdots,\Omega_N)=\prod_{i<j}\chi(r_{ij}).
\end{equation}
Expressing $\Psi$ in terms of chord distances { $r_{ij}$} regularizes the wave function and automatically takes care of the boundary condition when two atoms are on the opposite poles of the sphere ($\theta_{ij}=\pi$). For the Jastrow factor we choose $\chi(r)=K_0(2e^{-\gamma}r/a)e^{-r/b}$. With this choice $\Psi_0$ satisfies the Bethe-Peierls condition at short interparticle distances ($\theta_{ij}\approx r_{ij}\rightarrow 0$) as in the two-body case (see Sec.~\ref{Sec:2body}), and we control the system size by tuning the variational parameter $b$. 

In projection Monte Carlo methods, a first-order phase transition can be captured by imposing appropriate symmetry on the guiding wave functions. Probably, the most famous example of a zero-temperature quantum phase transition is that of the liquid-solid phase transition in $^4$He for which Monte Carlo calculations provide very good agreement with experiments~[\onlinecite{WhitlockCeperleyChesterKalos79}]. The distinguishing feature between the two phases is the order parameter, with the liquid phase having translational invariance, which is instead broken in the solid phase, where particles get localized near the lattice sites. On the variational level, close to the phase transition point, the variational energy has two minima as a function of the localization size. One corresponds to a solid (localization size smaller than the mean interparticle distance), while the other to a liquid (infinite localization size). The variational energy increases by going away from the minimum, so that each phase is stable against a weak perturbation, the phase with the lower energy being the ground state and the one with the higher energy being metastable. In contrast, in a second-order phase transition, only a single variational minimum exists, so that only one phase is stable at a time. 

In our case, the variational parameter $b$ describes the degree of localization of the state and allows one to discriminate between the two phases. For a sufficiently large number of particles ($N \gtrsim 20$), indeed, we observe the double-minimum structure, characteristic of a first-order phase transition, see Fig.~\ref{fig3}. The two minima are characterized by different values of $b$ and correspond, respectively, to the gas phase and to the localized soliton phase. The double minimum is observed in the variation energy in a narrow region of $x$ and the critical $x$ (when the two minima are degenerate) approaches the MF prediction $x\to -0.171$ as the number of particles is increased. 

DMC method is based on solving the Schrödinger equation in imaginary time and it allows one to find the ground-state properties in the limit of long projection time. The convergence is significantly improved by using a guiding wave function~(\ref{Jastrow}) with values of optimal $b$ obtained by minimizing the variational energy. The procedure is in principle exact and is independent of the choice of the guiding function, as long as all relevant configurations are allowed. We find that for $N=32$ and 128 Eq.~(\ref{Jastrow}), with the variationally optimal values of $b$, well describes both the gas and soliton solutions, and for these particle numbers, we do profit from importance sampling. However, since in the many-body configurational space, the ``gas'' guiding function to a large extent excludes the ``soliton'' region and vice versa, in practice, the diffusion process cannot account for the tunneling between the two phases. Quantitative description of the tunneling process in this case is an interesting future project going beyond the scope of this paper.

To mention a few technical details of our DMC scheme specific to the spherical geometry we find it convenient to work with particles' three-dimensional coordinates ${\bf r}_i$ rather than with their polar angles and azimuths. In particular, the chord distance evaluation in this case is much more straightforward. The drift for particle $i$ is governed by the gradient of $\Psi_0$ with respect to ${\bf r}_i$ projected to the sphere surface. For the Jastrow product (\ref{Jastrow}), this task reduces to calculating the projected gradient of $\chi$
\begin{equation}\label{Drift}
\bar\nabla_{{\bf r}_i}\chi(|{\bf r}_i-{\bf r}_j|)=\chi'(r_{ij})\frac{({\bf r}_i{\bf r}_j){\bf r}_i-{\bf r}_j}{r_{ij}}.
\end{equation}
Neglecting finite time step corrections, we realize the diffusion and drift in a three-dimensional manner (as if the atoms were not confined), then projecting back to the sphere surface. The local energy can be expressed in terms of derivatives of $\chi$ by using
\begin{equation}\label{LocalEn}
\hat{L}^2_{{\bf r}_i}\chi(|{\bf r}_i-{\bf r}_j|)=-\frac{1+{\bf r}_i{\bf r}_j}{2}\chi''(r_{ij})+\left(\frac{r_{ij}}{4}-\frac{{\bf r}_i{\bf r}_j}{r_{ij}}\right)\chi'(r_{ij}).
\end{equation}

Finally, we mention that another Monte Carlo method, path integral Monte Carlo, has recently been employed to study supersolidity of bosons on a sphere with soft-core and dipolar interactions at finite temperature~[\onlinecite{ciardi2023}].

\section{Conclusions}

In this paper, we address the problem of $N$ bosons on a spherical surface interacting via attractive zero-range potential and investigate how this system crosses over from the uniform gas state to the localized soliton state as one increases the attraction. Our main finding is that for sufficiently large $N$, we are essentially dealing with a first-order transition or, more precisely, a narrow avoided crossing between two states, which are significantly different in size and separated from each other by an energy barrier. On the other hand, for low values of $N$ the energy and rms size are smooth functions of the scattering length.

Our exact Monte Carlo calculations show that the change between these two regimes occurs when $N$ is roughly between $10$ and $20$. These relatively high values may be explained by a quantitatively weak geometry-induced breaking of the two-dimensional MF scaling invariance and, therefore, by a numerically small barrier. Indeed, in the MF approximation at $g=g_c$ the barrier equals $\Delta E_{\rm max}= 0.0336 N$ and we can compare it with the dipole mode frequency in the uniform phase $\epsilon_{1,1}=\sqrt{1-g_c/g_d}=0.262$. Here we have in mind that the dipole mode marks the initial ``direction'' for the system to tunnel toward the soliton side~[\onlinecite{Arovas}]. We see that for large $N$ the barrier is indeed much higher than the dipole frequency, the tunneling is suppressed, and the gas-soliton transition looks discontinuous (from the practical viewpoint). By contrast, when $\Delta E_{\rm max}\lesssim \epsilon_{1,1}$, the role of the barrier is not very important, and the crossover is smooth. Accordingly, the condition $\Delta E_{\rm max}=\epsilon_{1,1}$ leads to a characteristic atom number $N \approx 8$ where the crossover physics changes. This estimate is consistent with our Monte Carlo results.

We mention that the crossover scenario on a sphere is different from the previously solved model of attractive one-dimensional bosons on a ring where the gas-soliton transition is always continuous (no barrier), independent of $N$~[\onlinecite{Lincoln},\onlinecite{kanamoto2003}]. In perspective, it would therefore be interesting to study other geometries where the confinement can be used as a knob for controlling the crossover type and for elucidating the relative role of the space curvature, finite size, manifold topology, MF and beyond-MF terms. Being able to control the barrier and $N$ gives a way to prepare and probe superpositions of different macroscopic or mesoscopic quantum states.

\begin{acknowledgments}
We acknowledge support from ANR grant ``Droplets'' No.~ANR-19-CE30-0003-02 and from the EU Quantum Flagship (PASQuanS2.1, 101113690). 
G.E.A. acknowledges support by the Spanish Ministerio de Ciencia e Innovación (MCIN/AEI/10.13039/501100011033, grant PID2020-113565GB-C21), and by the Generalitat de Catalunya (grant 2021 SGR 01411). 
We also acknowledge Santander Supercomputacion support group at the University of Cantabria who provided access to the supercomputer Altamira Supercomputer at the Institute of Physics of Cantabria (IFCA-CSIC), member of the Spanish Supercomputing Network, for performing simulations/analyses.
ICFO group acknowledges support from:
European Research Council AdG NOQIA; 
MCIN/AEI (PGC2018-0910.13039/501100011033,  CEX2019-000910-S/10.13039/501100011033, Plan National FIDEUA PID2019-106901GB-I00, Plan National STAMEENA PID2022-139099NB, I00,project funded by MCIN/AEI/10.13039/501100011033 and by the “European Union NextGenerationEU/PRTR" (PRTR-C17.I1), FPI); QUANTERA MAQS PCI2019-111828-2);  QUANTERA DYNAMITE PCI2022-132919, QuantERA II Programme co-funded by European Union’s Horizon 2020 program under Grant Agreement No 101017733);
Ministry for Digital Transformation and of Civil Service of the Spanish Government through the QUANTUM ENIA project call - Quantum Spain project, and by the European Union through the Recovery, Transformation and Resilience Plan - NextGenerationEU within the framework of the Digital Spain 2026 Agenda;
Fundació Cellex;
Fundació Mir-Puig; 
Generalitat de Catalunya (European Social Fund FEDER and CERCA program, AGAUR Grant No. 2021 SGR 01452, QuantumCAT \ U16-011424, co-funded by ERDF Operational Program of Catalonia 2014-2020); 
Barcelona Supercomputing Center MareNostrum (FI-2023-1-0013); 
Funded by the European Union. Views and opinions expressed are however those of the author(s) only and do not necessarily reflect those of the European Union, European Commission, European Climate, Infrastructure and Environment Executive Agency (CINEA), or any other granting authority.  Neither the European Union nor any granting authority can be held responsible for them (EU Quantum Flagship PASQuanS2.1, 101113690, EU Horizon 2020 FET-OPEN OPTOlogic, Grant No 899794),  EU Horizon Europe Program (This project has received funding from the European Union’s Horizon Europe research and innovation program under grant agreement No 101080086 NeQSTGrant Agreement 101080086 — NeQST); 
ICFO Internal “QuantumGaudi” project; 
European Union’s Horizon 2020 program under the Marie Sklodowska-Curie grant agreement No 847648;  
“La Caixa” Junior Leaders fellowships, La Caixa” Foundation (ID 100010434): CF/BQ/PR23/11980043.

\end{acknowledgments}

\section*{Data Availability Statement}
The data that support the findings of this study are available from the authors upon reasonable request.

\section*{Author Declarations}
\subsection*{Conflict of interest}
The authors have no conflicts to disclose.

\appendix

\section{Derivation of Eq.~(\ref{EnUniformRen})}
\label{appendix}
A complementary manner of renormalizing the interaction is to expand the $N$-body energy in powers of $|g|\ll 1$ starting from the kinetic energy term in Eq.~(\ref{Ham}) as the unperturbed Hamiltonian. This standard perturbation theory works for finite $N$ and also predicts the nonpairwise energy contribution. For $N$ bosons on a unit sphere we obtain the ground-state energy up to terms of order $g^3$ in the form (see Ref.~[\onlinecite{PricoupenkoPetrov2021}])
\begin{equation}\label{EnUniformPert}
E_N=[g^{(1)}+g^{(2)}+g^{(3)}]\frac{N(N-1)}{2}+g_3\frac{N(N-1)(N-2)}{6},
\end{equation} 
where
\begin{equation}\label{g21}
g^{(1)}=V_{0 0}^{0 0}=\frac{g}{4\pi},
\end{equation}
\begin{equation}\label{g22}
g^{(2)}=-\sum_\nu\frac{|V_{0 \nu}^{0{\bar{\nu}}}|^2}{2\xi_\nu}=\left(\frac{g}{4\pi}\right)^2\left[\ln\frac{e^{1-2\gamma}}{l_c^2}+o(1)\right],
\end{equation}
\begin{equation}\label{g23}
\begin{aligned}
g^{(3)}=&\sum_\nu \sum_{\nu'}\frac{V_{0\nu'}^{0\bar{\nu}'}V_{\nu'\nu}^{\bar{\nu}'\bar{\nu}}V_{\nu 0}^{\bar{\nu} 0}}{4\xi_\nu\xi_{\nu'}}-V_{0 0}^{0 0}\sum_\nu\frac{|V_{0 \nu}^{0{\bar{\nu}}}|^2}{4\xi_\nu^2}\\
=&\left(\frac{g}{4\pi}\right)^3\left[\ln^2\frac{e^{1-2\gamma}}{l_c^2}-1+o(1)\right],
\end{aligned}
\end{equation}
and 
\begin{equation}\label{g33}
g_3=6\sum_\nu\frac{V_{0\nu}^{0\bar{\nu}}V_{\bar{\nu} 0}^{0 \bar{\nu}}V_{\nu 0}^{\bar{\nu} 0}}{4\xi_\nu^2}=6\left(\frac{g}{4\pi}\right)^3[1+o(1)].
\end{equation}
In Eqs.~(\ref{g22})-(\ref{g33}), $\nu=\{l,m\}$, $\nu'=\{l',m'\}$, $\bar{\nu}=\{l,-m\}$, $\bar{\nu}'=\{l',-m'\}$, $\sum_\nu=\sum_{l=1}^{l_c}\sum_{m=-l}^{l}$, and $\xi_\nu=l(l+1)/2$, where $l$ and $m$ correspond to the angular momentum and its projection for spherical harmonics. The quantities 
\begin{equation}
V_{{\bm \mu}{\bm \nu}}^{{\bm \zeta}{\bm \eta}} = g\int \! d \Omega \psi^*_{{\bm \zeta}} (\Omega) \psi^*_{{\bm \mu}}(\Omega) \psi_{{\bm \eta}}(\Omega) \psi_{{\bm \nu}}(\Omega),\label{Vequa}
\end{equation}
are the interaction matrix elements for two-body transitions from single-particle states $\mu$ and $\zeta$ to $\nu$ and $\eta$ and the corresponding wave functions are spherical harmonics. The last terms in Eqs.~(\ref{g22})-(\ref{g33}) are the final results obtained by summing over $\nu$ and $\nu'$. The summation is facilitated by the fact that $V_{0\nu}^{0\bar{\nu}}=V_{\bar{\nu} 0}^{0 \bar{\nu}}=V_{\nu 0}^{\bar{\nu} 0}=g/(4\pi)$. The more general matrix element $V_{\nu'\nu}^{\bar{\nu}'\bar{\nu}}$ in the first sum in Eq.~(\ref{g23}) is not necessary to calculate explicitly. The result in this case is obtained by summing over the projections $m$ and $m'$ and using the addition theorem for spherical harmonics. Finally, to arrive at Eq.~(\ref{EnUniformRen}) of the main text, we rewrite the expansion (\ref{EnUniformPert}) in powers of $g_r$ by using the relation $1/g=1/g_r+\ln(e^{-\gamma+1/2}/l_c)/(2\pi)$, which follows from { Eqs.~(\ref{gkappa}) and (\ref{gr})}. Note that the cutoff dependence drops out. 

{

\section{Derivation of Eq.~(\ref{BvsRresult})}
\label{Sec:BMFflat}

From Eqs.~(\ref{sigmaN}) and (\ref{BN}) obtained by Hammer and Son~[\onlinecite{hammer2004}] one can conclude that the product $|B_N|\langle r_{ij}^2\rangle=\exp[o(N)]$. This does not exclude that this product may be a power of $N$ since the term $o(N)$ can, in principle, be logarithmic. Here, by developing the beyond-MF description of the flat soliton we show that its size and energy are related by Eq.~(\ref{BvsRresult}), i.e., the product $|B_N|\langle r_{ij}^2\rangle$ tends to a universal constant.

The planar version of the Hamiltonian Eq.~(\ref{Ham}) reads
\begin{equation}\label{Hamplane}
\hat{H}_{\rm plane} = \frac{1}{2}\int d^2\rho (-\hat{\psi}_{\bf \rho}^{\dagger} \nabla^2_{\bf \rho} \hat{\psi}_{\bf \rho} + g \hat{\psi}_{\bf \rho}^{\dagger} \hat{\psi}_{\bf \rho}^{\dagger} \hat{\psi}_{\bf \rho} \hat{\psi}_{\bf \rho}), 
\end{equation}
and the corresponding mean-field Gross-Pitaevskii energy functional is
\begin{equation}\label{GPenergy}
B_{\rm MF}(\Psi)=\frac{1}{2}\int d^2\rho [|\nabla_{\bf \rho}\Psi({\bf \rho})|^2+g|\Psi({\bf \rho})|^4].
\end{equation} 
Minimizing Eq.~(\ref{GPenergy}) with the normalization constraint $\int d^2\rho |\Psi({\bf \rho})|^2=N$ leads to stationary solutions only for $g=g_c=-4\pi/(N\ln r)$~[\onlinecite{chiao1964}]. The corresponding wave functions form a family of self-similar states parametrized by the size $\sigma$,
\begin{equation}\label{PsiR}
\Psi_\sigma({\bf \rho})=\sqrt{\frac{N\ln r}{8\pi \sigma^2}}f(\rho/\sigma),
\end{equation}
where $f(\rho)$ is a nodeless solution of 
\begin{equation}\label{NLSE}
f''(\rho)+f'(\rho)/\rho-f(\rho)+f^3(\rho)=0.
\end{equation}
The function $f(\rho)$ is found numerically, it has a bell shape with exponentially decaying large-$\rho$ asymptote~[\onlinecite{chiao1964},\onlinecite{hammer2004}]. The normalization integral, the kinetic energy, and the interaction energy corresponding to the solution $\Psi_\sigma$ can be found from the identities
\begin{equation}\label{Integrals}
\int d\rho \rho f^2(\rho)=\int d\rho \rho[f'(\rho)]^2=\frac{1}{2}\int d\rho \rho f^4(\rho)=\frac{4}{\ln r}.
\end{equation}
The rms separation of atoms in the Townes soliton is related to the size parameter $\sigma$ by
\begin{equation}\label{rho2}
\sqrt{\langle r_{ij}^2\rangle} =\sigma\sqrt{2\int d\rho \rho^3 f^2(\rho)/\int d\rho \rho f^2(\rho)}=1.541\sigma,
\end{equation}
where we use
\begin{equation}\label{rho2bare}
\int d\rho \rho^3 f^2(\rho)=2.211.
\end{equation}
For future reference, we also provide the integral
\begin{equation}\label{fprime2bare}
\int d\rho \rho^3 [f'(\rho)]^2=2.968.
\end{equation}

One can check that if $g=g_c$ the total energy $B_{\rm MF}(\Psi_\sigma)=0$ for any $\sigma$. This result, in a more general form, follows from the stationarity condition~[\onlinecite{Vlasov}]. Although the kinetic energy and the interaction energy in Eq.~(\ref{GPenergy}) cancel each other, they are separately of order $gN^2/\sigma^2\sim N/\sigma^2$, defining the MF energy scale. The leading beyond-MF contribution, which we denote $B_{\rm BMF}$, is smaller by the factor $|g|\sim 1/N\ll 1$. The shape of $\Psi$ is fixed by the dominant MF forces. The beyond-MF term is too weak to induce significant changes in the shape but it breaks the degeneracy associated with the parameter $\sigma$. The optimal $\sigma=\sigma_N$ is determined by minimizing the energy functional including the beyond-MF term $B_{\rm BMF}(\Psi_\sigma)$. Calculating this quantity is a complicated task, which requires diagonalizing the Bogoliubov Hamiltonian for fluctuations around the stationary condensate solution $\Psi_\sigma$. In this manner one should, in principle, obtain the size $\sigma_N$ (and, therefore, $\sqrt{\langle r_{ij}^2\rangle}$) and the energy $B_N$ up to the preexponential factors [we are speaking about the terms $o(N)$ in Eqs.~(\ref{sigmaN}) and (\ref{BN})], eventually arriving at Eq.~(\ref{BvsRresult}). We will now show that determining the product $B_N \sigma_N^2$ does not require these complicated calculations; it is sufficient to know the scaling properties of $B_{\rm BMF}(\Psi_\sigma)$. 

In contrast to Eq.~(\ref{GPenergy}), the Hamiltonian Eq.~(\ref{Hamplane}) is not scaling invariant because of the (implicit) cutoff $\kappa$ in the interaction. We can nevertheless use the following scaling property: a scaling transformation (change of coordinates by a factor $\Lambda$) of a many-body eigenstate of the Hamiltonian Eq.~(\ref{Hamplane}) corresponding to a cutoff $\kappa$ gives an eigenstate of the same Hamiltonian, but with $1/\kappa$ rescaled by $\Lambda$. This rescaling of the cutoff momentum is equivalent to rescaling the scattering length $a$ by $\Lambda$ [see Eq.~(\ref{gkappa})]. The corresponding energy gets rescaled by $1/\Lambda^2$. This means that the beyond-MF term $B_{\rm BMF}(\Psi_\sigma,\kappa)$ calculated with the cutoff $\kappa$ for the condensate wave function $\Psi_\sigma$ and the analogous quantity $B_{\rm BMF}(\Psi_{\sigma'},\kappa')$ calculated with the cutoff $\kappa'=\kappa \sigma/\sigma'$ for $\Psi_{\sigma'}$ are related by 
\begin{equation}\label{EBMFrescaling}
B_{\rm BMF}(\Psi_\sigma,\kappa)=B_{\rm BMF}(\Psi_{\sigma'},\kappa')(\sigma'/\sigma)^2.
\end{equation}
Let us now assume that we know $B_{\rm BMF}(\Psi_{\sigma'},\kappa)$ for a certain reference size $\sigma'$ and we want to calculate $B_{\rm BMF}(\Psi_{\sigma},\kappa)$, i.e., we want to know how the beyond-MF term scales with $\sigma$ for a given fixed $\kappa$. In view of Eq.~(\ref{EBMFrescaling}), it is sufficient to calculate $B_{\rm BMF}(\Psi_{\sigma'},\kappa')$ which we write as $B_{\rm BMF}(\Psi_{\sigma'},\kappa')=B_{\rm BMF}(\Psi_{\sigma'},\kappa)+\Delta B$. Here, $\Delta B=B_{\rm BMF}(\Psi_{\sigma'},\kappa')-B_{\rm BMF}(\Psi_{\sigma'},\kappa)$ is the difference between the beyond-MF terms calculated for the same condensate wave function $\Psi_{\sigma'}$ but different cutoff momenta. This is a local term since it takes into account excitations with de Broglie wave lengths between $1/\kappa$ and $1/\kappa'$, both assumed to be much smaller than the soliton size $\sigma'$. When calculating $\Delta B$, one can thus use the Bogoliubov theory for homogeneous condensates (and then average over the density) or just simply note that changing $\kappa\rightarrow \kappa'$ corresponds to a renormalization of the coupling constant given by the second-order Born integral. The result is [by $o(1)$ we mean terms $\ll 1$ in the asymptotic limit $|g|\sim 1/N\rightarrow 0$]
\begin{widetext}
\begin{equation}\label{DiffEkappa}
\Delta B= -\int d^2 \rho \frac{|\Psi_{\sigma'}(\rho)|^4}{2}\int_\kappa^{\kappa'}\frac{2\pi kd k}{(2\pi)^2}\frac{g^2}{k^2}+o(1)=-\frac{N^2g^2\ln r}{(4\pi \sigma')^2}\ln\frac{\kappa'}{\kappa}+o(1)=-\frac{N^2g^2\ln r}{(4\pi \sigma')^2}\ln\frac{\sigma}{\sigma'}+o(1).
\end{equation}

Using the fact that $B_{\rm MF}(\Psi_\sigma)=(g-g_c)\int d^2\rho|\Psi_\sigma({\bf \rho})|^4/2$ and Eqs.~(\ref{EBMFrescaling}) and (\ref{DiffEkappa}), we can write the energy of a soliton of size $\sigma$ up to the leading beyond-MF correction as
\begin{equation}\label{MFplusBMF}
B(\Psi_\sigma)=\frac{N^2(g-g_c)\ln r}{8\pi \sigma^2}+B_{\rm BMF}(\Psi_{\sigma'},\kappa)\left(\frac{\sigma'}{\sigma}\right)^2-\frac{N^2g^2\ln r}{(4\pi \sigma)^2}\ln\frac{\sigma}{\sigma'}.
\end{equation}
\end{widetext}
Minimizing Eq.~(\ref{MFplusBMF}) with respect to $\sigma$ leads to the result 
\begin{equation}\label{AlmostThere}
B_N\sigma_N^2=-\frac{N^2g^2\ln r}{32\pi^2}.
\end{equation} 
The point is that $B_N$ and $\sigma_N$ separately require calculating $B_{\rm BMF}$, whereas the product Eq.~(\ref{AlmostThere}) does not.
Our theory implies that $|g-g_c|\ll |g|$ so that the first term on the right-hand side of Eq.~(\ref{MFplusBMF}) is consistent with the beyond-MF energy scale. Not to exceed the accuracy we should then also replace $g$ by $g_c$ in the last term in Eq.~(\ref{MFplusBMF}) as well as in Eq.~(\ref{AlmostThere}). Equation~(\ref{BvsRresult}) then follows from Eqs.~(\ref{AlmostThere}) and (\ref{rho2}). 

To complete the proof of Eq.~(\ref{BvsRresult}), we still have one technical point to address. Even when $a$ is fixed, we can choose different $g$ to model the interaction, provided that we tune $\kappa$ in accordance with Eq.~(\ref{gkappa}). This freedom is limited by two constraints: $|g-g_c|\ll |g|$, discussed above, and $\kappa \gg 1/\sigma_N$. The second constraint allows us to neglect finite-range effects and guarantees that our theory can be used to describe zero-range interactions. In fact, for the accuracy claimed in Eq.~(\ref{DiffEkappa}), we need $\kappa$ to exceed $1/\sigma_N$ by at least one power of $N$. The obvious choice $g=g_c$ satisfies the first constraint. It corresponds to $\kappa_c=2a^{-1}\exp[N(\ln r)/2-\gamma]=\sigma_N^{-1}\exp[o(N)]$ as follows from Eqs.~(\ref{gkappa}), (\ref{sigmaN}), and (\ref{rho2}). Since $o(N)$ is not necessarily asymptotically large, $\kappa_c$ may not be larger than $\sigma_N^{-1}$ and we cannot ensure that the interaction is sufficiently short ranged. Let us then choose $\kappa = \sigma_N^{-1}N$ and show that the corresponding $g$ satisfies $|g-g_c|\ll |g|$. To this end we write
\begin{equation}\label{gthroughgc}
g=\frac{2\pi}{\ln(2e^{-\gamma}/\kappa_c a)+\ln(\kappa_c/\kappa)}\approx g_c-\frac{g_c^2}{2\pi}\ln\frac{\kappa_c}{\kappa}
\end{equation}
and note that $\kappa_c/\kappa=\exp[o(N)]$, which means $|g-g_c|=g^2o(N)\ll |g|$ as we need. This point completes the proof of Eq.~(\ref{BvsRresult}).

\section{Derivation of Eq.~(\ref{DeltaEsphereGen})}

Let us find how the energy of a nearly flat soliton is shifted under the influence of the sphere. The idea is to perturbatively calculate the curvature induced MF energy shift assuming that $\sigma\ll 1$ and using the flat-soliton wave function Eq.~(\ref{PsiR}) as the unperturbed solution. Consider a cylindrically symmetric function $\Psi(\rho)$ where $\rho=2\sin(\theta/2)$ is the chord distance between the soliton center, assumed to be at the north pole, and the point on the sphere with polar angle $\theta$. In this case we have $d\Omega= 2\pi \sin \theta d\theta = 2\pi \rho d\rho$, and the operator $\hat{L}^2$ defined after Eq.~(\ref{Ham}) can be written as
\begin{equation}
\hat{L}^2=-\frac{\partial^2}{\partial {\rho}^2}-\frac{1}{\rho}\frac{\partial}{\partial {\rho}}+\frac{\rho^2}{4}\left(\frac{\partial^2}{\partial {\rho}^2}+\frac{3}{\rho}\frac{\partial}{\partial {\rho}}\right).
\end{equation}
The MF energy functional on the sphere can thus formally be written as 
\begin{equation}\label{GPenergySphere}
E_{\rm MF}(\Psi)=B_{\rm MF}(\Psi)+\int d^2\rho \frac{\rho^2}{8}\Psi^*(\rho)\left(\frac{\partial^2}{\partial {\rho}^2}+\frac{3}{\rho}\frac{\partial}{\partial {\rho}}\right)\Psi(\rho),
\end{equation} 
where $B_{\rm MF}(\Psi)$ is given by Eq.~(\ref{GPenergy}), and the domain of $\rho$ is restricted to be between 0 and 2 with a proper boundary condition for $\Psi$ at $\rho=2$. However, the corresponding finite-size corrections are exponentially small for $\sigma\ll 1$, and we can extend the domain up to $\rho=\infty$. The leading-order curvature induced energy shift is thus triggered by the integral in Eq.~(\ref{GPenergySphere}) and can easily be calculated for $\Psi=\Psi_\sigma$. Using Eq.~(\ref{fprime2bare}), we obtain 
\begin{equation}\label{MFshiftCurve}
\Delta E_{\rm sphere}(\Psi_\sigma)=-\frac{N\ln r}{32}\int [f'(\rho)]^2\rho^3 d\rho=-0.199 N,
\end{equation}
which is Eq.~(\ref{DeltaEsphereGen}). Although Eq.~(\ref{MFshiftCurve}) is formally a MF correction, it is by a factor of $\sigma^2$ weaker than the MF energy scale $~N/\sigma^2$. It becomes comparable to the beyond-MF scale if $\sigma^2\sim 1/N$, which is exactly what happens at the transition point, as we mention in the main text. Note that Eq.~(\ref{MFshiftCurve}) is independent of $\sigma$ and thus it only shifts the energy, the optimal $\sigma$ still being determined by the beyond-MF energy term Eq.~(\ref{MFplusBMF}) derived in the flat-surface approximation. 
}

\end{document}